\documentclass[         %
aps,                    
pre,                    
showplace,               
nofootinbib,            
showkeys,               %
preprintnumbers,        %
floatfix]               
{revtex4}               
\usepackage{graphicx,longtable,multirow}
\usepackage[dvips]{epsfig}

\begin{document}
\title{Spectra and Symmetry in Nuclear Pairing}
\author{        A.~B. Balantekin}
\email{         baha@physics.wisc.edu}
\author{        J. H. de Jesus}
\email{         jhjesus@physics.wisc.edu}
\author{        Y. Pehlivan}
\email{         yamac@physics.wisc.edu}
\affiliation{  Department of Physics, University of Wisconsin \\
               Madison, Wisconsin 53706 USA }
\date{\today}
\begin{abstract}
We apply the algebraic Bethe ansatz technique to the nuclear
pairing problem with orbit dependent coupling constants and
degenerate single particle energy levels. We find the exact
energies and eigenstates. We show that for a given shell, there
are degeneracies between the states corresponding to less and more
than half full shell. We also provide a technique to solve the
equations of Bethe ansatz.
\end{abstract}
\medskip
 \preprint{} \maketitle

\section{Introduction}

Pairing is known to play an important role in the quantum
many-body problems. In particular, pairing force is a key
ingredient of the residual interaction between nucleons in a
nucleus. An algebraic approach to pairing was given by Richardson
some time ago \cite{rich}. Richardson's formalism was rather
complex and was not widely used in nuclear physics. For a limited
class of seniority-conserving pairing interactions, the quasispin
formalism of Kerman \cite{kerman1} was used to treat a large
number of cases. In this article, we wish to explore an approach
related to Richardson's formalism which can be reduced to the
quasispin formalism in the appropriate limit.

In the nuclear shell model, the nucleus is pictured as a system of
fermions moving in a central field with well defined single
particle energy levels $\epsilon_j$, arising from spin-orbit
interactions.  We consider nucleons at time-reversed states $|j \>
m \rangle$ and $(-1)^{(j-m)}|j \> -m\rangle$, interacting with a
pairing force described by the Hamiltonian
\begin{equation}\label{1}
\hat{H}=\sum_{jm} \epsilon_j a^\dagger_{j\>m} a_{j\>m} -
|G|\sum_{jj'}c_{jj'} \hat{S}^+_j \hat{S}^-_{j'}.
\end{equation}
Here, $a^\dagger_{j\>m}$ and $a_{j\>m}$ are the creation and
annihilation operators for nucleons in level $j$, respectively,
and $c_{jj'}$ is the strength of the pairing interaction between
levels $j$ and $j'$. The quasispin operators $\hat{S}^+_j$ and
$\hat{S}^-_{j}$ are given by
\begin{eqnarray}\label{2}
\hat{S}^+_j&=&\sum_{m>0} (-1)^{(j-m)} a^\dagger_{j\>m}a^\dagger_{j\>-m},\\
\hat{S}^-_j&=&\sum_{m>0} (-1)^{(j-m)} a_{j\>-m}a_{j\>m}.\nonumber
\end{eqnarray}
These operators create or destroy a single pair of nucleons in the
time reversed states on level $j$. If we also define
\begin{equation}\label{3}
\hat{S}^0_j=\frac{1}{2}\sum_{m>0}
\left(a^\dagger_{j\>m}a_{j\>m}+a^\dagger_{j\>-m}a_{j\>-m}-1
\right),
\end{equation}
together with the operators in Eqs. (\ref{2}) we have a set of
orthogonal $SU(2)$ algebras
\begin{equation}\label{4}
[\hat{S}^+_j,\hat{S}^-_{j'}]=2\delta_{jj'}\hat{S}^0_j \ \ \ \ \ \
[\hat{S}^0_j,\hat{S}^\pm_{j'}]=\pm\delta_{jj'}\hat{S}^\pm_j.
\end{equation}
Note that $\hat{S}^0_j$ can also be written as
\begin{equation}\label{5}
\hat{S}^0_j=\hat{N}_{j}-\frac{1}{2}\Omega_j.
\end{equation}
Here $\Omega_j=j+\frac{1}{2}$ is the maximum number of pairs that
can occupy level $j$ and $\hat{N}_j$ is the pair number operator
level $j$ which is given by
\begin{equation}\label{6}
\hat{N}_j=\frac{1}{2}\sum_{m>0}
\left(a^\dagger_{j\>m}a_{j\>m}+a^\dagger_{j\>-m}a_{j\>-m}\right).
\end{equation}
$\hat{N}_j$ takes values from zero to $\Omega_j$ and thus
$\hat{S}^0_j$ takes values from $-\frac{1}{2}\Omega_j$ to
$\frac{1}{2}\Omega_j$. Consequently, the $SU(2)$ spanned by
$\hat{S}^\pm_j$ and $\hat{S}^0_{j}$ live in the spin
$\frac{1}{2}\Omega_j$ representation.

When the pairing strength is separable ($c_{jj'}=c^*_jc_{j'}$),
the Hamiltonian given in Eq. (\ref{1}) takes the form
\begin{equation}\label{7}
\hat{H}= \sum_{jm} \epsilon_j a^\dagger_{j\>m} a_{j\>m} -
|G|\sum_{jj'}c^*_jc_{j'} \hat{S}^+_j \hat{S}^-_{j'}.
\end{equation}
Furthermore, if we assume that the energy levels are degenerate,
the first term is a constant for a given number of pairs because
the Hamiltonian is number conserving. Ignoring this term, we
obtain
\begin{equation}\label{8}
\hat{H}=- |G|\sum_{jj'}c^*_jc_{j'} \hat{S}^+_j \hat{S}^-_{j'}.
\end{equation}
Defining the operators
\begin{equation}\label{9}
\hat{S}^+(0)=\sum_j c^*_j\hat{S}^+_j \ \ \ \ \mbox{and} \ \ \ \ \
\hat{S}^-(0)=\sum_j c_j\hat{S}^-_j,
\end{equation}
the Hamiltonian in Eq. (\ref{8}) can be written as
\begin{equation}\label{10}
\hat{H}=-|G|\hat{S}^+(0)\hat{S}^-(0).
\end{equation}
The reason that we choose to label the operators in Eq. (\ref{9})
as $\hat{S}^\pm(0)$ will become clear in what follows. Physically,
the operator $\hat{S}^+(0)$ creates a single fermion pair and
$c^*_j$ can be viewed as the probability amplitude that this pair
is created at level $j$. This interpretation implies, however,
that these constants are normalized:
\begin{equation}\label{11}
\sum_j |c_j|^2=1.
\end{equation}
Note that a state with no pairs, here denoted by $|0\rangle$, is
the lowest weight state for all the $SU(2)$ algebras corresponding
to different levels. In other words, it obeys
\begin{equation}\label{12}
\hat{S}^-_j|0\rangle=0 \ \ \ \mbox{and} \ \ \
\hat{S}^0_j|0\rangle=-\frac{\Omega_j}{2}|0\rangle \ \ \ \
\mbox{for every $j$}.
\end{equation}
Therefore, this state is annihilated by the Hamiltonian in Eq.
(\ref{8})
\begin{equation}\label{13}
\hat{H}|0\rangle=0.
\end{equation}
The state which represents the full shell, denoted by
$|\bar{0}\rangle$ in this paper, is the highest weight state of
all $SU(2)$ algebras corresponding to different levels. It obeys
\begin{equation}\label{14}
\hat{S}^+_j|\bar{0}\rangle=0 \ \ \ \mbox{and} \ \ \
\hat{S}^0_j|\bar{0}\rangle=\frac{\Omega_j}{2}|\bar{0}\rangle \ \ \
\ \mbox{for every $j$}.
\end{equation}
The state $|\bar{0}\rangle$ is also an eigenstate of the
Hamiltonian
\begin{equation}\label{15}
\hat{H}|\bar{0}\rangle=E_1|\bar{0}\rangle.
\end{equation}
Its energy is given by
\begin{equation}\label{16}
E_1=-|G|\sum_j \Omega_j |c_j|^2.
\end{equation}

Unlike the fully occupied or empty shell, it is considerably more
difficult to calculate the eigenstates and the energies when the
shell is partially occupied. In a study of states with generalized
seniority zero, Talmi  \cite{talmi} considered states of two
particles coupled to angular momentum zero. He showed that under
certain assumptions, a state of the form
\begin{equation}\label{17}
\hat{S}^+(0)|0\rangle=\sum_j c^*_j\hat{S}^+_j |0\rangle
\end{equation}
is an eigenstate of a class of Hamiltonians including that in Eq.
(\ref{8}). It is fairly straightforward to show that the
Hamiltonian in Eq. (\ref{8}) satisfies
\begin{equation}\label{18}
\hat{H}\hat{S}^+(0)|0\rangle=E_1\hat{S}^+(0)|0\rangle
\end{equation}
where $E_1$ is given by Eq. (\ref{16}). We see that the state in
Eq. (\ref{17}) which has one pair of nucleons has the same energy
as the full shell state $|\bar{0}\rangle$. Obviously, the state
given in Eq. (\ref{17}) is not the only eigenstate for one pair of
nucleons. For example, for two levels with total angular momenta
$j_1$ and $j_2$, it is straightforward to show that the state
\begin{equation}\label{19}
\left( \frac{c_{j_2}}{\Omega_{j_1}} \hat{S}^+_{j_1} -
\frac{c_{j_1}}{\Omega_{j_2}} \hat{S}^+_{j_2} \right) |0 \rangle,
\end{equation}
which is orthogonal to the state (\ref{17}), is also an eigenstate
of the Hamiltonian in Eq. (\ref{8}).

One may ask if there is a systematic way to derive these states.
The authors of Ref. \cite{Pan:1997rw} came up with an elegant
solution to this question. They calculated the energy eigenvalues
and eigenstates of the Hamiltonian given in Eq. (\ref{8}) using
the operators
\begin{equation}\label{20}
\hat{S}^+(x)=\sum_j\frac{c^*_j}{1-|c_j|^2x}\hat{S}^+_j \ \ \ \ \
\mbox{and} \ \ \ \ \
\hat{S}^-(x)=\sum_j\frac{c_j}{1-|c_j|^2x}\hat{S}^-_j.
\end{equation}
Here, $x$ is a parameter which can be real or complex. In this
technique, one starts from the empty shell $|0\rangle$ and
constructs Bethe ansatz states using the operators $S^+(x)$.
Substitution of these states into the energy equation yields Bethe
ansatz equations (BAE) which determine the values of the
parameters $x$. Note that the operators given in Eq. (\ref{9}) are
special cases of the operators defined in Eq. (\ref{20}) for
$x=0$. The authors of Ref. \cite{Pan:1997rw} successfully adopted
the procedure outlined above and obtained the energy eigenstates
and eigenvalues of the Hamiltonian given in Eq. (\ref{8}). Their
proof, however, relies on a Laurent expansion of the operators
$\hat{S}^\pm(x)$ around $x=0$. Later, they use an analytic
continuation argument to vindicate the validity of their results
in the entire complex plane except some singular points.

In this paper, we give an alternative, algebraic proof which does
not impose the analyticity condition and valid in general for any
complex values of the constants $c_j$. We also show that the same
technique works for the pairs of holes. In other words, the Bethe
ansatz states can be constructed with the operators $S^-(x)$
acting the fully occupied shell $|\bar{0}\rangle$. As we will
show, one obtains the same Bethe ansatz equations and the same
energies for states with $N$ nucleon pairs and $N-1$ hole pairs
(which is equivalent to $N_{max}+1-N$ particle pairs). Here
$N_{max}$ is the maximum number of pairs that can occupy the shell
in question and $N\leq N_{max}/2$. The symmetry between the states
$|\bar{0}\rangle$ and $\hat{S}^+(0)|0\rangle$ pointed out in Eqs.
(\ref{15}) and (\ref{18}) is a special case of this symmetry for
$N=1$.

Here, we also provide techniques for solving the resulting Bethe
ansatz equations. One should point out that it is also possible to
solve the pairing problem numerically in the quasispin basis
\cite{Volya:2000ne}. Our approach is complementary to this
numerical approach.

This paper is organized as follows: In Section II, we outline the
Bethe ansatz formalism and give the eigenstates and eigenvalues of
the pairing Hamiltonian. In this section, we also compare our
results with the quasispin limit of the problem. In Section III,
we consider the special case of a shell with two levels. We show
that, the problem of solving the equations of Bethe ansatz with
two levels for zero energy eigenstates can be transformed into a
problem of finding the roots of an hypergeometric polynomial. In
Section IV, we compare our results with the available data in the
literature. We present our conclusions in Section V.

\section{Eigenstates and Eigenvalues of the Nuclear Pairing Hamiltonian}

We begin by introducing the operator
\begin{equation}\label{21}
\hat{K}^0(x)=\sum_j\frac{1}{1/|c_j|^2-x}\hat{S}_j^0
\end{equation}
in addition to the operators in Eq. (\ref{20}). It is
straightforward to show that the following commutators are
satisfied by $\hat{S}^+(x)$, $\hat{S}^-(x)$ and
$\hat{K}^0(x)$\footnote{Note that the operators
$\hat{S}^\pm(x)$and $\hat{K}^0(x)$ can be written in terms of the
rational Gaudin algebra generators which appear in the context of
the reduced BCS pairing \cite{Richardson}. If we take the
occupation amplitudes $c_j$ to be real, then the relation is
simply
\begin{eqnarray*}
\hat{S}^\pm(x)=\frac{\hat{J}^\pm(\sqrt{x})+\hat{J}^\pm(-\sqrt{x})}{2},
\ \ \ \ \ \ \ \ \ \
\hat{K}^0(x)=\frac{\hat{J}^0(\sqrt{x})-\hat{J}^0(-\sqrt{x})}{2\sqrt{x}}
\end{eqnarray*}
where
\begin{eqnarray*}
\hat{J}^a(\lambda)=\sum_j\frac{\hat{S}^a_j}{1/c_j-\lambda}
\end{eqnarray*}
are the rational Gaudin algebra generators for $a=0,\pm$.}:
\begin{eqnarray}\label{22}
[\hat{S}^+(x),\hat{S}^-(0)]&=&[\hat{S}^+(0),\hat{S}^-(x)] =
2K^0(x),
\\ \label{23} [\hat{K}^0(x),\hat{S}^\pm(y)] &=&\pm
\frac{\hat{S}^\pm(x)-\hat{S}^\pm(y)}{x-y}.
\end{eqnarray}
When $x=y$, the limit has to be taken on both sides of Eq.
(\ref{23}) as $y\to x$. The states $|0\rangle$ and
$|\bar{0}\rangle$ defined with Eqs. (\ref{12}) and (\ref{14}),
respectively, are eigenstates of $\hat{K}^0(x)$:
\begin{equation}\label{24}
\hat{K}^0(x)|0\rangle =
-\left(\sum_j\frac{\Omega_j/2}{1/|c_j|^2-x}\right)|0\rangle ,\ \ \
\ \ \ \ \hat{K}^0(x)|\bar{0}\rangle =
\left(\sum_j\frac{\Omega_j/2}{1/|c_j|^2-x}\right)|\bar{0}\rangle .
\end{equation}

Below, we construct the eigenstates of the Hamiltonian $\hat{H}$
using the algebraic Bethe ansatz formalism. We first consider the
eigenstates with one pair of nucleons in order to demonstrate the
technique. Then we discuss the general case with more than one
pair.

\subsection{Eigenstates With One Pair of Nucleons}

Let us first form a Bethe ansatz state as follows:
\begin{equation}\label{25}
\hat{S}^+(x^{(1)})|0\rangle.
\end{equation}
We denoted our variable by $x^{(1)}$ in order to emphasize that
this state has one pair of nucleons. Using the form of the
Hamiltonian given in Eq. (\ref{10}), together with Eq. (\ref{13})
and Eqs. (\ref{22})-(\ref{24}), we can show that the Hamiltonian
acting on this state gives
\begin{equation}\label{26}
\hat{H}\hat{S}^+(x^{(1)})|0\rangle=-2|G|
\left(\sum_j\frac{\Omega_j/2}{1/|c_j|^2-x^{(1)}}\right)\hat{S}^+(0)|0\rangle.
\end{equation}
Setting $x^{(1)}=0$ in the state (\ref{25}), one obtains the state
in Eq. (\ref{17}). One can see from Eq. (\ref{26}) that the
corresponding energy is as in Eq. (\ref{16}).

 Alternatively, we can choose $x^{(1)}$ as a solution of
\begin{equation}\label{27}
\sum_j \frac{-\Omega_j/2}{1/|c_j|^2-x^{(1)}}=0.
\end{equation}
Then Eq. (\ref{26}) becomes
\begin{equation}\label{28}
\hat{H}\hat{S}^+(x^{(1)})|0\rangle=0.
\end{equation}
In other words, for the values of $x^{(1)}$ satisfying Eq.
(\ref{27}), the state given in Eq. (\ref{25}) is an eigenstate of
$\hat{H}$ with zero energy. Here Eq. (\ref{27}) is our one-pair
Bethe ansatz equation because it determines the value of
$x^{(1)}$.

The eigenstates with one pair of nucleons discussed here can also
be seen in Table \ref{Table1}. We see that when there is only one
pair of nucleons occupying the shell, the Hamiltonian $\hat{H}$ of
Eq. (\ref{8}) has only one eigenstate with nonzero energy which is
$\hat{S}^+(0)|0\rangle$. All other eigenstates are orthogonal to
this state and have zero energy. In fact, the BAE (\ref{27}) can
be seen as this orthogonality relation. In many cases, the BAE in
Eq. (\ref{27}) has more than one solution and each solution gives
us an eigenstate in the form of Eq. (\ref{25}) with zero energy.
As discussed in Section III, in the presence of two levels, Eq.
(\ref{27}) has only one solution and this solution gives the state
(\ref{19}). In the presence of, for example, three levels, Eq.
(\ref{27}) has two distinct solutions and therefore we have two
zero energy states. These states can be seen in Table
\ref{Table5}.

\subsection{Eigenstates With More Than One Pair of Nucleons}

The eigenstates of the Hamiltonian in Eq. (\ref{8}) with more than
one nucleon pair can also be written in the Bethe ansatz. Some of
these states have zero energy, like the state in Eq. (\ref{25}),
and some of them have nonzero energy, like the state in Eq.
(\ref{17}). We examine these states below and the results of this
section are summarized in Table \ref{Table1}.

\begin{table}
\begin{tabular}{|c|c|c|c|}
\hline &  & & \\ %
 {\bf \# of pairs} & {\bf Bethe Ansatz Equations}
 & {\bf State} & {\bf Energy$/\left(-|G|\right)$} \\ & & & \\
\hline \multicolumn{4}{|c|}{Less than half full or half full shell} \\ \hline & & & \\ %
\multirow{4}{*}{$1$} & $\displaystyle \sum_j
\frac{-\Omega_j/2}{1/|c_j|^2-x^{(1)}}=0$ & $\displaystyle
\hat{S}^+(x^{(1)})|0\rangle$ & 0
\\ & & & \\ \cline{2-4} & & & \\ & No BAE
& $\displaystyle \hat{S}^+(0)|0\rangle$  & $\displaystyle \sum_j \Omega_j |c_j|^2$ \\ & & & \\
\hline  & & & \\ %
\multirow{5}{*}{$N$} & $\displaystyle \sum_j
\frac{-\Omega_j/2}{1/c_j^2-x^{(N)}_m}=\sum_{k=1(k\neq m)}^N
\frac{1}{x^{(N)}_m-x^{(N)}_k}$ & $\displaystyle
\hat{S}^+(x^{(N)}_1)\hat{S}^+(x^{(N)}_2) \dots
\hat{S}^+(x^{(N)}_N)|0\rangle$ & 0 \\ & {\footnotesize For
$m=1,2,\dots,N$} & &
\\ \cline{2-4} & & & \\ & $\displaystyle \sum_j \frac{-\Omega_j/2}{1/|c_j|^2-z^{(N)}_m}
=\frac{1}{z^{(N)}_m}+\sum_{k=1(k\neq m)}^{N-1}
\frac{1}{z^{(N)}_m-z^{(N)}_k}$ & $\displaystyle
\hat{S}^+(0)\hat{S}^+(z^{(N)}_1) \dots
\hat{S}^+(z^{(N)}_{N-1})|0\rangle$ & $\displaystyle \sum_j
\Omega_j |c_j|^2-\sum_{k=1}^{N-1} \frac{2}{z^{(N)}_k}$ \\ &
{\footnotesize For $m=1,2,\dots,N-1$} & &  \\
\hline \multicolumn{4}{|c|}{More than half full shell} \\ \hline & & & \\ %
$\displaystyle N_{max}+1-N$ & $\displaystyle \sum_j
\frac{-\Omega_j/2}{1/|c_j|^2-z^{(N)}_m}
=\frac{1}{z^{(N)}_m}+\sum_{k=1(k\neq m)}^{N-1}
\frac{1}{z^{(N)}_m-z^{(N)}_k}$ & $\displaystyle
\hat{S}^-(z_1^{(N)})\hat{S}^-(z_2^{(N)})\dots\hat{S}^-(z_{N-1}^{(N)})|\bar{0}\rangle$
& $\displaystyle \sum_j \Omega_j |c_j|^2-\sum_{k=1}^{N-1}
\frac{2}{z^{(N)}_k}$ \\ &
{\footnotesize For $m=1,2,\dots,N-1$} & &  \\
\hline  & & & \\ %
$\displaystyle N_{max}$ & No BAE & $\displaystyle |\bar{0}\rangle$
& $\displaystyle \sum_j \Omega_j |c_j|^2$ \\ & & & \\
\hline
\end{tabular}
\caption{Summary of the energy eigenvalues and the eigenstates of
the pairing Hamiltonian. Here, $N_{max}$ denotes the maximum
number of pairs which can occupy the shell and $\displaystyle
2\leq N\leq N_{max}/2$. For shells which are at most half full,
$N$ corresponds to the number of pairs as can be seen in the upper
part of the first column. For shells which are more than half
full, $N$ is related to the number of pairs as indicated in the
lower part of the first column. Note that for more than half full
shells there are no zero energy states.}\label{Table1}
\end{table}

\subsubsection{The Eigenstates With Non-zero Energy}

Using the form of the Hamiltonian given in Eq. (\ref{10}),
together with Eq. (\ref{13}) and (\ref{22})-(\ref{24}), it is
possible to show that the state
\begin{equation}\label{29}
\hat{S}^+(0)\hat{S}^+(z^{(N)}_1) \dots
\hat{S}^+(z^{(N)}_{N-1})|0\rangle
\end{equation}
is an eigenstate of the Hamiltonian if the parameters $z^{(N)}_k$
obey the following Bethe ansatz equations\footnote{If we assume
that all the coefficients $c_j$ are real, Eqs. (\ref{30}) and
(\ref{32}) are the same as the equations obtained by Pan
{\textit{et al}} in Ref. \cite{Pan:1997rw}. Note that, instead of
$z^{(N)}_m$ they write their equations in terms of the new
variables
\begin{eqnarray*}
y_m=\frac{z^{(N)}_m}{\displaystyle\sum_{k=1}^{N-1}\frac{1}{z^{(N)}_k}}
\end{eqnarray*}
and they also use the definitions
\begin{eqnarray*}
\Lambda_1=-\frac{1}{2}\sum_j \Omega_j c_j^2 \ \ \ \ \ \mbox{and} \
\ \ \ \ \alpha=\sum_{k=1}^{N-1}\frac{1}{z^{(N)}_k}.
\end{eqnarray*}
Substituting these definitions into Eqs. (\ref{30}) and (\ref{32})
reproduces their result.} (see the Appendix)
\begin{equation}\label{30}
\sum_j \frac{-\Omega_j/2}{1/|c_j|^2-z^{(N)}_m}
=\frac{1}{z^{(N)}_m}+\sum_{k=1(k\neq m)}^{N-1}
\frac{1}{z^{(N)}_m-z^{(N)}_k}, \ \ \ \ \ \ \mbox{for every} \ \
m=1,2,\dots N-1.
\end{equation}
The state in Eq. (\ref{29}) has $N$ pairs of nucleons and this is
the reason we use the superscript $(N)$. Here, we are assuming
that $2\leq N \leq N_{max}/2$. The case for one nucleon is already
examined in Section II.A and if the shell is more than half full,
we choose to work with hole pairs instead of particle pairs.
Therefore the state (\ref{30}) represents a shell which is at most
half full. The details of this calculation can be found in the
Appendix. Here, we would like to emphasize that the BAE's
(\ref{30}) are a set of $N-1$ coupled equations in $N-1$
variables. The parameters $z^{(N)}_m$ are all different from one
another and they are also different from zero.

Using the same parameters $z^{(N)}_m$ that appear in Eq.
(\ref{29}), we now form the state
\begin{equation}\label{31}
\hat{S}^-(z_1^{(N)})\hat{S}^-(z_2^{(N)})\dots\hat{S}^-(z_{N-1}^{(N)})|\bar{0}\rangle.
\end{equation}
This state has $N_{max}+1-N$ pairs of nucleons because it starts
from the full shell with $N_{max}$ pairs of nucleons, represented
by $|\bar{0}\rangle$ and then destroys $N-1$ pairs. Since we
assumed that $2\leq N \leq N_{max}/2$, this state represents a
shell which is more than half full. In the Appendix, we show that
if the parameters $z^{(N)}_k$ obey the equations of Bethe ansatz
given in Eqs. (\ref{30}), then the state (\ref{31}) is also an
eigenstate of the Hamiltonian like the state (\ref{29}). We can
also show that the states (\ref{29}) and (\ref{31}) have the same
energy which is given in terms of the variables $z^{(N)}_k$ as
follows:
\begin{equation}\label{32}
E_N =-|G|\left(\sum_j \Omega_j |c_j|^2-\sum_{k=1}^{N-1}
\frac{2}{z^{(N)}_k}\right).
\end{equation}

At this point, we would like to remark that the solutions of the
BAE given in Eqs. (\ref{30}) may be complex. Nevertheless, since
the complex solutions always come in conjugate pairs, the energy
in Eq. (\ref{32}) is always real.

In principle, Eqs. (\ref{30}) may have more than one solution in
which case each solution gives us two eigenstates. One should
substitute each solution in Eqs. (\ref{29}) and (\ref{31}) in
order to find corresponding eigenstates and then in Eq. (\ref{32})
in order to find their energy.

The state in Eq. (\ref{29}) can be thought as the generalization
of the state in Eq. (\ref{17}) found by Talmi \cite{talmi}. For
$N=2$, the state (\ref{29}) becomes
\begin{equation}\label{33}
\hat{S}^+(0)\hat{S}^+(z^{(2)}_1)|0\rangle
\end{equation}
and the state (\ref{31}) becomes
\begin{equation}\label{34}
\hat{S}^-(z^{(2)}_1)|\bar{0}\rangle.
\end{equation}
The state in Eq. (\ref{33}) has $2$ nucleon pairs whereas the
state in Eq. (\ref{34}) has $N_{max}-1$ nucleon pairs. We have
only one unknown variable, $z^{(2)}_1$, and the BAE we must solve
to determine it can be found by substituting $N=2$ in Eqs.
(\ref{30}). Note that the sum in the last term of Eqs. (\ref{30})
does not contain any terms for $N=2$ so that the Bethe ansatz
equation is
\begin{equation}\label{35}
\sum_j \frac{-\Omega_j/2}{1/|c_j|^2-z^{(2)}_1}
=\frac{1}{z^{(2)}_1}.
\end{equation}
From  Eq. (\ref{32}), we find the energy of the states in Eqs.
(\ref{33}) and (\ref{34}) to be
\begin{equation}\label{36}
E_2=-|G|\left(\sum_j \Omega_j |c_j|^2-\frac{2}{z^{(2)}_1}\right).
\end{equation}
By solving Eq. (\ref{35}), one determines the values of
$z^{(2)}_1$ and then substitutes each solution in the states in
Eqs. (\ref{33}) and (\ref{34}) in order to find the corresponding
eigenstates and then in Eq. (\ref{36}) in order to find the
energies of these states.

We would like to note that although we call the states in Eqs.
(\ref{29}) and (\ref{31}) nonzero energy states, in principle, the
energy calculated from Eq. (\ref{32}) may turn out to be zero for
some specific values of the parameters $\Omega_j$ and $c_j$.
Nevertheless, for generic values of the parameters of the problem,
the energies of the states described here are different from zero
as opposed to the states we examine below which are annihilated by
the Hamiltonian and therefore have identically zero energy.

\subsubsection{The Eigenstates with Zero Energy}

In the Appendix, we show that a state of the form
\begin{equation}\label{37}
\hat{S}^+(x^{(N)}_1)\hat{S}^+(x^{(N)}_2) \dots
\hat{S}^+(x^{(N)}_N)|0\rangle
\end{equation}
is annihilated by the Hamiltonian $\hat{H}$ of Eq. (\ref{8}), if
the parameters $x^{(N)}_k$ satisfy the following Bethe ansatz
equations:
\begin{equation}\label{38}
\sum_j \frac{-\Omega_j/2}{1/|c_j|^2-x^{(N)}_m}=\sum_{k=1(k\neq
m)}^N \frac{1}{x^{(N)}_m-x^{(N)}_k}, \ \ \ \ \ \ \mbox{for every}
\ \ \ m=1,2,\dots,N.
\end{equation}
The state in Eq. (\ref{37}) has $N$ nucleon pairs where $2\leq N
\leq N_{max}/2$. Here, the parameters $x^{(N)}_k$ may be real or
complex, but they are all different from one another.

We would like to emphasize that the Bethe ansatz equations
(\ref{38}) may have more than one solution. In this case, each
solution gives us an eigenstate in the form of Eq. (\ref{37}) with
zero energy. It may also be the case that the BAE in Eqs.
(\ref{38}) do not admit any solutions for some $N$. This is not
surprising given the fact that it may not always be possible to
find a zero energy configuration for any number of pairs.

\subsection{Quasispin Limit}

The results obtained in Sections II.A and II.B reveal a symmetry
between the nonzero energy spectra of the pairing Hamiltonian with
$N$ pairs and $N_{max}+1-N$ pairs of nucleons. Here $N_{max}$ is
the total number of pairs that can occupy the shell and $N\leq
N_{max}/2$. This means that for a given shell, there are
degeneracies between the eigenstates of the nuclei corresponding
to less and more than half full shell. Here we examine the
quasispin limit of the problem to gain insight about this
symmetry. We also show that zero energy states do not obey this
symmetry because there are no zero energy states for the nuclei
corresponding to more than half full shell\footnote{Obviously, one
can always shift the spectra of the nuclei so that the ground
state has zero energy. But here what we mean by a zero energy
state is a state annihilated by the pairing Hamiltonian given in
Eq. (\ref{8}). For example, in Figure \ref{fig1}, they correspond
to the highest energy state for each nuclei. Therefore, if we
shift each spectra so that the ground state has zero energy, it is
the highest excited state which is missing in the spectra of the
nuclei corresponding to more than half full shell. This can also
be seen in Figure \ref{fig2}.}.

Let us consider a shell consisting of $d$ levels $j_1$, $j_2$,
$\dots$, $j_d$ with maximum occupancies $\Omega_{j_1}$,
$\Omega_{j_2}$, $\dots$, $\Omega_{j_d}$. If we assume that the
occupation amplitudes for these levels are all equal to one
another, i.e., $c_{j_1}=c_{j_2}=\dots=c_{j_d}$, then the pairing
Hamiltonian of Eq. (\ref{8}) is proportional to
\begin{equation}\label{44}
Q^+Q^- =
\overrightarrow{Q}\cdot\overrightarrow{Q}-Q^0\left(Q^0-1\right)
\end{equation}
where
\begin{equation}\label{sum}
\overrightarrow{Q}=\overrightarrow{S}_{j_1}+
\overrightarrow{S}_{j_2}+\dots+\overrightarrow{S}_{j_d}
\end{equation}
is the sum of the quasispins corresponding to different levels.
Since the quasispin quantum number corresponding to level $j$ is
$\Omega_j$, Eq. (\ref{sum}) corresponds to the addition of the
angular momenta $\Omega_{j_1}/2$, $\Omega_{j_2}/2$, $\dots$,
$\Omega_{j_d}/2$. In this limit, the eigenstates of the pairing
Hamiltonian are $|J\> M\rangle$, where $J$ is the total angular
momentum. The energies are given by
\begin{equation}\label{45}
Q^+Q^-|J\> M\rangle %
=\left[J(J+1)-M(M-1)\right]|J\> M\rangle.
\end{equation}
Clearly, the lowest weight states $|J\> -J\rangle$ have zero
energy. These are the quasispin limit of the zero energy states
given in Eq. (\ref{37}). The other states which have nonzero
energies can be written as $|J\>\tilde{M}\rangle$ and
$|J\>1-\tilde{M}\rangle$ if we define $-J+1\leq \tilde{M} \leq 0$.
These states are the quasispin limit of the states given in Eqs.
(\ref{29}) and (\ref{31}), respectively. To see this, first note
that from Eq. (\ref{5}) we find
\begin{equation}\label{48}
Q^0=\sum_j \hat{S}^0_{j}=\hat{N}_{tot} - \frac{N_{max}}{2}
\end{equation}
where $\hat{N}_{tot}$ is the total pair number operator. We can
use this equation to find the number of pairs in a state
$|J\>M\rangle$ as
\begin{equation}\label{NUMBER}
\hat{N}_{tot}|J\>M\rangle
=\left(M+\frac{N_{max}}{2}\right)|J\>M\rangle.
\end{equation}
This equation tells us that if the state $|J\>\tilde{M}\rangle$
has $N$ pairs of nucleons, then the state $|J\>1-\tilde{M}\rangle$
has $N_{max}+1-N$ pairs of nucleons where $N\leq N_{max}/2$. We
also see from Eq. (\ref{45}) that the states
$|J\>\tilde{M}\rangle$ and $|J\>1-\tilde{M}\rangle$ have the same
energy. This is consistent with the results found above, i.e.,
that the nonzero energy eigenstates of the pairing Hamiltonian
with $N$ pairs and $N_{max}+1-N$ pairs have the same energy.
Although this symmetry is trivial in the quasispin limit, we have
proved that it is valid for generic values of occupation
amplitudes $c_j$ as well.

The quasispin limit of the problem also tells us about the zero
energy states of the pairing Hamiltonian for a given number of
pairs. In general, it is difficult to tell if a zero energy
configuration exists for a given number of pairs just by looking
at the equations of Bethe ansatz (\ref{38}). These equations are
not always guaranteed to have solutions leading to nontrivial
states\footnote{The state $|\psi\rangle=0$ is the trivial solution
of $\hat{H}|\psi\rangle=0$. Sometimes, even if the Bethe ansatz
equations do have solutions, the corresponding state may be equal
to zero.}. Nevertheless, Eq. (\ref{10}) tells us that the zero
energy states of the pairing Hamiltonian are the states
annihilated by $\hat{S}^-(0)$ defined in Eq. (\ref{9}). In the
quasispin limit, $\hat{S}^-(0)$ is proportional to $Q^-$ and the
zero energy states become the lowest weight states
$|J\>-J\rangle$. Therefore, we can restrict ourselves to the
quasispin limit as long as the discussion is limited to the
existence of zero energy states of the pairing Hamiltonian with a
given number of pairs. In this case, Eq. (\ref{NUMBER}) tells us
that the number of pairs in a zero energy state $|J\>-J\rangle$ is
$N_{max}/2-J$. Since $J\geq 0$, we conclude that zero energy
states do not exist for a more than half full shell.

In many cases, we can come to an even stronger conclusion on the
existence of the zero energy states by examining the quasispin
limit. For example, in the presence of two levels $j_1$ and $j_2$,
quasispin limit corresponds to the addition of the two angular
momenta $\Omega_{j_1}/2$ and $\Omega_{j_2}/2$. Assuming that
$\Omega_{j_2}>\Omega_{j_1}$, the sum of the two angular momenta
yields the following set of orthogonal representations:
\begin{equation}\label{41}
J=\frac{\Omega_{j_2}+\Omega_{j_1}}{2},\frac{\Omega_{j_1}+\Omega_{j_2}}{2}-1,
\frac{\Omega_{j_1}+\Omega_{j_2}}{2}-2\dots,
\frac{\Omega_{j_2}-\Omega_{j_1}}{2}.
\end{equation}
We have a zero energy state $|J\>-J\rangle$ for each one of them.
From Eq. (\ref{NUMBER}) we see that the lowest weight state
$|J\>-J\rangle$ has $N_{max}/2-J$ nucleon pairs. In this special
case $N_{max}=\Omega_{j_2}+\Omega_{j_1}$. Using this, we find that
the zero energy states exists only for
$N=0,1,2,\dots,\Omega_{j_1}$. Since we set $\Omega_1$ as the
smaller of $\Omega_1$ and $\Omega_2$ and since there is no
degeneracy in Eq. (\ref{41}), we conclude that, in the presence of
two levels, one and only one zero energy state exists for each
number of pairs until there are enough pairs to fill one of the
levels. After that, no zero energy state exists.

\section{Exact Solution for Two Nuclear Energy Levels}

In the previous section, we showed that in a system of two nuclear
levels, zero energy states of the pairing Hamiltonian exist only
for small number of pairs, i.e.,  when the number of pairs is not
more than enough to fill any one of the levels. Here, we consider
the Bethe ansatz equations which are to be solved in order to find
the states annihilated by the pairing Hamiltonian (i.e. Eqs.
(\ref{27}) and (\ref{38})) for two levels and for arbitrary values
of the occupation probabilities $|c_{j_1}|^2$ and $|c_{j_1}|^2$.
We give analytical solutions of these equations in the form of the
roots of some hypergeometric polynomials. The method we use here
is adopted from Ref. \cite{Balantekin:2004yf}.

The Bethe ansatz equations for states annihilated by $\hat{H}$
were given by Eq. (\ref{27}) for pair of nucleons and by Eqs.
(\ref{38}) for more than one pair of nucleons. When there are only
two levels, these equations can be written as
\begin{equation}\label{49}
\frac{-\Omega_{j_1}/2}{1/|c_{j_1}|^2-x_i^{(N)}}
+\frac{-\Omega_{j_2}/2}{1/|c_{j_2}|^2-x_i^{(N)}}=\sum_{k=1(k\neq
i)}^{N}\frac{1}{x_{i}^{(N)}-x_{k}^{(N)}},
\end{equation}
for $i=1,2,\dots,N$. These equations are to be satisfied for every
$x_i^{(N)}$ so that we have a system of $N$ coupled nonlinear
equations. If $N=1$, then the sum on the right hand side of Eq.
(\ref{49}) is identically zero and this covers the one pair case
described by Eq. (\ref{27}). The $N$ variables
$\{x_1^{(N)},x_2^{(N)},\dots,x_N^{(N)}\}$ which satisfy these
equations give us the state with $N$ pairs when we substitute them
in Eq. (\ref{37}) whose special case for $N=1$ is Eq. (\ref{27}).

Let us begin by introducing the variables $\eta_i^{(N)}$ which are
related to $x_i^{(N)}$ with the linear transformation
\begin{equation}\label{50}
x_{i}^{(N)}=\frac{1}{|c_{j_2}|^2}+\eta_i^{(N)}
\left(\frac{1}{|c_{j_1}|^2}-\frac{1}{|c_{j_2}|^2}\right).
\end{equation}
Here, we assumed that $|c_{j_1}|^2 \neq |c_{j_2}|^2$ (if the
occupation probabilities $|c_{j_1}|^2$ and $|c_{j_2}|^2$ are equal
to each other, then the problem is reduced to the quasispin limit
as described in the previous section and can easily be solved).
When we write the BAE (\ref{49}) in terms of the new variables
introduced in Eq. (\ref{50}), we find
\begin{equation}\label{51}
\sum_{k=1(k\neq i)}^{N}\frac{1}{\eta_{i}^{(N)}-\eta_{k}^{(N)}}
-\frac{\Omega_{j_2}/2}{\eta_i^{(N)}}+\frac{\Omega_{j_1}/2}{1-\eta_i^{(N)}}=0,
\end{equation}
for $i=1,2,\dots,N$. This way, the dependence of the BAE on the
occupation probabilities $|c_{j_1}|^2$ and $|c_{j_2}|^2$
disappears.

In Ref. \cite{stiel}, Stieltjes had shown that the $N^{th}$ order
polynomial
\begin{equation}\label{52}
p_N(z) = \prod_{i =1}^{N} (z - \eta_i^{(N)})
\end{equation}
whose roots obey Eqs. (\ref{51}), satisfies the hypergeometric
differential equation
\begin{equation}\label{53}
z(1-z) p_{N}^{\prime \prime }(z)
+\left[-\Omega_{j_2}+\left(\Omega_{j_1}+\Omega
_{j_2}\right)z\right] p_{N}^{\prime }(z) +
N\left(N-\Omega_{j_1}-\Omega_{j_2}-1\right) p_N(z)=0.
\end{equation}
Therefore, finding the $N^{th}$ order polynomial solution of this
differential equation and then finding the roots of this
polynomial gives us the solution of the BAE. Since the
differential equation is of second order, it has two solutions.
One solution is an hypergeometric polynomial of order $N$ and the
other is another hypergeometric polynomial of order
$\Omega_1+\Omega_2+1-N$. We are only interested in the $N^{th}$
order polynomial solution because of Eq. (\ref{52}). As before, we
can choose $\Omega_{j_1}<\Omega_{j_2}$, without loss of
generality. We also set $N\leq\Omega_{j_1}$ since we have shown
that states annihilated by $\hat{H}$ does not exist otherwise.
Under these conditions, the $N^{th}$ order polynomial solution of
the differential equation in Eq. (\ref{53}) is the following
hypergeometric function \cite{Erdelyi}:
\begin{equation}\label{54}
p_{N}(z)
=F(-N,N-\Omega_{j_1}-\Omega_{j_2}-1;-\Omega_{j_2};z)=\sum_{k=0}^{N}\frac{(-N)_k
\left(N-\Omega_{j_1}-\Omega_{j_2}-1\right)_k}{\left(-\Omega_{j_2}\right)_k(k!)}z^k.
\end{equation}
Here, $(a)_k$ is the Pochhammer number which is defined as
\begin{equation}\label{55}
(a)_k=\frac{\Gamma(a+k)}{\Gamma(a)}=a\left(a+1\right)\left(a+2\right)\dots\left(a+k-1\right)
\end{equation}
for $k>1$. If $k=0$, then $(a)_0=1$. Pochhammer numbers have the
property that $(-k)_{k+1}=0$. But since $N<\Omega_{j_2}$, the
Pochhammer number $\left(-\Omega_{j_2}\right)_k$ in the
denominator of Eq. (\ref{54}) never vanishes.

This procedure reduces the problem of solving the BAE's (\ref{49})
for $N$ pairs to a problem of finding the roots of an $N^{th}$
order hypergeometric polynomial. Once the roots of the polynomial
(\ref{54}) are found, then Eq. (\ref{50}) can be used to find the
variables $x_i^{(N)}$. We then substitute $x_i^{(N)}$ in the state
(\ref{37}) in order to find corresponding zero energy eigenstates.
We illustrate the technique for $N=1$ and $N=2$ below.

For $N=1$, the hypergeometric function given in Eq. (\ref{54}) is
\begin{equation}\label{56}
F(-1,-\Omega_{j_1}-\Omega_{j_2};-\Omega_{j_2};z)
=1-\frac{\Omega_{j_1}+\Omega_{j_2}}{\Omega_{j_2}}z.
\end{equation}
This is a first order polynomial and its root is
\begin{equation}\label{57}
\eta_1^{(1)}=\frac{\Omega_{j_2}}{\Omega_{j_1}+\Omega_{j_2}}.
\end{equation}
When substituted in Eq. (\ref{50}), we obtain
\begin{equation}\label{58}
x^{(1)}=\frac{1}{|c_{j_2}|^2}+\frac{\Omega_{j_2}}{\Omega_{j_1}+\Omega_{j_2}}
\left(\frac{1}{|c_{j_1}|^2}-\frac{1}{|c_{j_2}|^2}\right).
\end{equation}
This gives us the state given in Table \ref{Table2} with $E=0$
where we omitted a normalization constant. In Table \ref{Table2},
we also give the eigenstate of the Hamiltonian $\hat{H}$ with
nonzero energy for completeness which is the state (\ref{17}).

\begin{table}
\begin{tabular}{|c|c|}
\hline
 {\bf Energy/$\left(-|G|\right)$} & {\bf State} \\
\hline & \\ %
$0$ &
$\left(-\frac{c_{j_2}}{\Omega_{j_1}}\hat{S}_{j_1}^++\frac{c_{j_1}}{\Omega_{j_2}}
\hat{S}_{j_2}^+\right)|0\rangle$ \\ & \\
\hline  & \\ %
$\Omega_{j_1} |c_{j_1}|^2+\Omega_{j_2} |c_{j_2}|^2$ & $\left(
c^*_{j_1}\hat{S}^+_{j_1} +
c^*_{j_2}\hat{S}^+_{j_2} \right)|0 \rangle$ \\ & \\
\hline
\end{tabular}
\caption{Energies and eigenstates for two levels and one pair of
nucleons.}\label{Table2}
\end{table}

\begin{table}
\begin{tabular}{|c|c|}
\hline
   {\bf Energy/$\left(-|G|\right)$} & {\bf State} \\
\hline & \\ %
$0$ %
&$\left(\frac{c^2_{j_2}}{\Omega_{j_1}\left(\Omega_{j_1}-1\right)}
\hat{S}_{j_1}^+\hat{S}_{j_1}^+
-\frac{2c_{j_1}c_{j_2}}{\Omega_{j_1}\Omega_{j_2}}
\hat{S}_{j_1}^+\hat{S}_{j_2}^+
+\frac{c^2_{j_1}}{\Omega_{j_2}\left(\Omega_{j_2}-1\right)}\hat{S}_{j_2}^+
\hat{S}_{j_2}^+\right)|0\rangle$ \\ & \\

\hline & \\ %
$\begin{array}{l}\frac{1}{2}\left(3\Omega_{j_1}-2\right)|c_{j_1}|^2+\\
\frac{1}{2}\left(3\Omega_{j_2}-2\right)|c_{j_2}|^2
-\frac{1}{2}\Lambda
\end{array}$
&%
$\left(c_{j_1}^{\ast}\hat{S}_{j_1}^++c_{j_2}^{\ast}\hat{S}_{j_2}^+\right)
\left[\begin{array}{l}-\frac{c_{j_1}^{\ast}}{2\Omega_{j_1}}
\left(\left(\Omega_{j_1}+\Omega_{j_2}\right)|c_{j_2}|^2
+\left(\Omega_{j_1}-2\right)\left(|c_{j_2}|^2-|c_{j_1}|^2\right)
+\Lambda\right) \hat{S}_{j_1}^+  \\
+\frac{c_{j_2}^{\ast}}{2\Omega_{j_2}}\left(\left(\Omega_{j_1}+\Omega_{j_2}\right)
|c_{j_1}|^2+\left(\Omega_{j_2}-2\right) \left(
|c_{j_1}|^2-|c_{j_2}|^2\right)
+\Lambda \right) \hat{S}_{j_2}^+%
\end{array}\right]|0\rangle$
\\ & \\

\hline & \\
$\begin{array}{l}\frac{1}{2}\left(3\Omega_{j_1}-2\right)|c_{j_1}|^2+\\
\frac{1}{2}\left(3\Omega_{j_2}-2\right)|c_{j_2}|^2+\frac{1}{2}\Lambda
\end{array}$
&%
$\left(c_{j_1}^{\ast}\hat{S}_{j_1}^++c_{j_2}^{\ast}\hat{S}_{j_2}^+\right)
\left[\begin{array}{l}-\frac{c_{j_1}^{\ast}}{2\Omega_{j_1}}
\left(\left(\Omega_{j_1}+\Omega_{j_2}\right)|c_{j_2}|^2
+\left(\Omega_{j_1}-2\right)\left(|c_{j_2}|^2-|c_{j_1}|^2\right)
-\Lambda\right) \hat{S}_{j_1}^+  \\
+\frac{c_{j_2}^{\ast}}{2\Omega_{j_2}}\left(\left(\Omega_{j_1}+\Omega_{j_2}\right)
|c_{j_1}|^2+\left(\Omega_{j_2}-2\right) \left(
|c_{j_1}|^2-|c_{j_2}|^2\right)
-\Lambda \right) \hat{S}_{j_2}^+%
\end{array}\right]|0\rangle$
\\ & \\
\hline
\end{tabular}
\caption{Energies and eigenstates for two levels and two pairs of
nucleons. Here, $\Lambda$ is the separation between the two
nonzero energies and it is given by Eq. (\ref{62}).}
\label{Table3}
\end{table}

For $N=2$, the hypergeometric polynomial in Eq. (\ref{54}) becomes
\begin{equation}\label{59}
F(-2,1-\Omega_{j_1}-\Omega_{j_2};-\Omega_{j_2};z)
=1-\frac{2\left(\Omega_{j_1}+\Omega_{j_2}-1\right)}{\Omega_{j_2}}z
+\frac{\left(\Omega_{j_1}+\Omega_{j_2}-1\right)\left(\Omega_{j_1}+\Omega_{j_2}-2\right)}
{\Omega_{j_2}\left(\Omega_{j_2}-1\right)}z^2.
\end{equation}
This is a second order polynomial whose roots are
\begin{equation}\label{60}
\eta_{1,2}^{(2)}=\frac{1}{\Omega_{j_1}+\Omega_{j_2}-2}\left(\left(
\Omega_{j_2}-1\right)\pm\sqrt{1-\frac{\Omega_{j_1}\Omega_{j_2}}
{\Omega_{j_1}+\Omega_{j_2}-1}}\right).
\end{equation}
Substituting these roots in Eq. (\ref{50}) we obtain the variables
\begin{equation}\label{61}
x_{1,2}^{(2)}=\frac{1}{|c_{j_2}|^2}+\frac{1}{\Omega_{j_1}+\Omega_{j_2}-2}\left(\left(
\Omega_{j_2}-1\right)\pm\sqrt{1-\frac{\Omega_{j_1}\Omega_{j_2}}
{\Omega_{j_1}+\Omega_{j_2}-1}}\right)
\left(\frac{1}{|c_{j_1}|^2}-\frac{1}{|c_{j_2}|^2}\right).
\end{equation}
When substituted in Eq. (\ref{37}), this solution yields the state
given in Table \ref{Table3} with $E=0$. The nonzero energy states
of the Hamiltonian $\hat{H}$ are also shown in Table \ref{Table3}.
These are found by solving the nonzero energy Bethe ansatz
equation (\ref{35}) for two levels $j_1$ and $j_2$. There are two
solutions of Eq. (\ref{35}) in the presence of two levels. Each
solution gives an eigenstate in the form of Eq. (\ref{33}). Their
energies, calculated using Eq. (\ref{36}), are also given in Table
\ref{Table3}. The two nonzero energies are separated by $\Lambda$
which is given by
\begin{equation}\label{62}
\Lambda=\sqrt{\left(\left(\Omega_{j_1}-2\right)
|c_{j_1}|^2+\left(\Omega_{j_2}-2\right) |c_{j_2}|^2\right)
^2+8\left(\Omega_{j_1}+\Omega_{j_2}-2\right)
|c_{j_1}|^2|c_{j_2}|^2}.
\end{equation}

The solutions of Eq. (\ref{35}) also give us eigenstates with
$N_{max}-1$ pairs in the form of Eq. (\ref{34}) as explained in
Section II. These states are given in Table \ref{Table4} and they
have the same energies as those in Table \ref{Table3}.


\begin{table}
\begin{tabular}{|c|c|}
\hline
 {\bf Energy/$\left(-|G|\right)$} & {\bf State} \\
\hline & \\ %
$\begin{array}{l}\frac{1}{2}\left(3\Omega_{j_1}-2\right)|c_{j_1}|^2+\\
\frac{1}{2}\left(3\Omega_{j_2}-2\right)|c_{j_2}|^2
-\frac{1}{2}\Lambda
\end{array}$
&%
$\left[\begin{array}{l}-\frac{c_{j_1}^{\ast}}{2\Omega_{j_1}}
\left(\left(\Omega_{j_1}+\Omega_{j_2}\right)|c_{j_2}|^2
+\left(\Omega_{j_1}-2\right)\left(|c_{j_2}|^2-|c_{j_1}|^2\right)
+\Lambda\right) \hat{S}_{j_1}^-  \\
+\frac{c_{j_2}^{\ast}}{2\Omega_{j_2}}\left(\left(\Omega_{j_1}+\Omega_{j_2}\right)
|c_{j_1}|^2+\left(\Omega_{j_2}-2\right) \left(
|c_{j_1}|^2-|c_{j_2}|^2\right)
+\Lambda \right) \hat{S}_{j_2}^-%
\end{array}\right]|\bar{0}\rangle$ \\ & \\ %
\hline & \\
$\begin{array}{l}\frac{1}{2}\left(3\Omega_{j_1}-2\right)|c_{j_1}|^2+\\
\frac{1}{2}\left(3\Omega_{j_2}-2\right)|c_{j_2}|^2+\frac{1}{2}\Lambda
\end{array}$
&%
$ \left[\begin{array}{l}-\frac{c_{j_1}^{\ast}}{2\Omega_{j_1}}
\left(\left(\Omega_{j_1}+\Omega_{j_2}\right)|c_{j_2}|^2
+\left(\Omega_{j_1}-2\right)\left(|c_{j_2}|^2-|c_{j_1}|^2\right)
-\Lambda\right) \hat{S}_{j_1}^-  \\
+\frac{c_{j_2}^{\ast}}{2\Omega_{j_2}}\left(\left(\Omega_{j_1}+\Omega_{j_2}\right)
|c_{j_1}|^2+\left(\Omega_{j_2}-2\right) \left(
|c_{j_1}|^2-|c_{j_2}|^2\right)
-\Lambda \right) \hat{S}_{j_2}^-%
\end{array}\right]|\bar{0}\rangle$
\\ & \\
\hline
\end{tabular}
\caption{Energies and eigenstates for two levels and $N_{max}-1$
pairs of nucleons. Here $\Lambda$ is the separation between the
energies and it is given by Eq. (\ref{62}).}\label{Table4}
\end{table}

\section{Results}

\begin{figure}[t]
  \centering
  \includegraphics{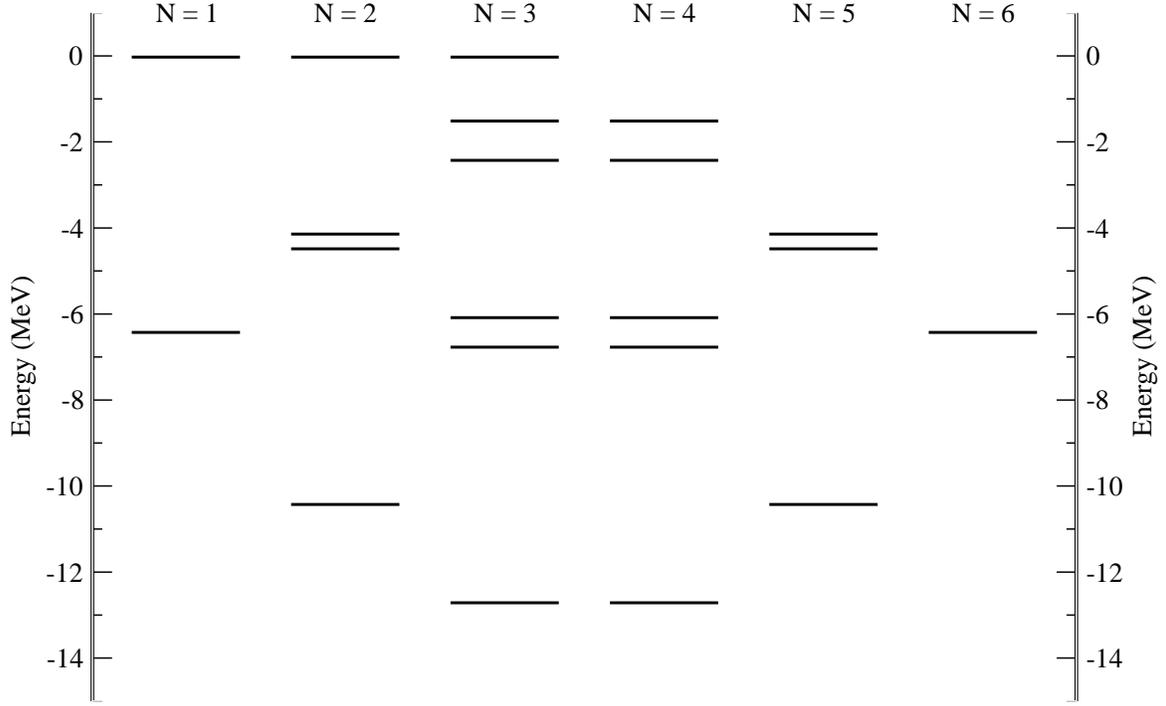} 
  \caption{Energy eigenvalues for $N=1-6$ pairs of nucleons in the first nuclear
   $sd$ shell. The coefficients $c_j$ and the pairing strength $|G|$ are those
   from \cite{Pan:1997rw}. Note the spectral symmetry discussed in the text and also
   the absence of the zero energy states for the nuclei corresponding to more than half
   full shell.}
  \label{fig1}
\end{figure}

In the work of Pan {\textit{et al}} ~\cite{Pan:1997rw}, an example
was provided where the pairing spectra was obtained for one, two
and three pairs of nucleons which are allowed to occupy the first
nuclear $sd$ shell. With the formalism developed in this paper, we
can easily complete that picture by including the spectra for
four, five and six pairs of nucleons. We do so, by considering the
vacuum consists of six pairs of nucleons fully occupying the
sub-shells $0d_{5/2}$, $0d_{3/2}$ and $1s_{1/2}$.  Then, we
construct a state with one pair of nucleon-holes, which
corresponds to the state with five pairs of nucleons and the state
with two pairs of nucleon-holes which corresponds to the state
with four pairs of nucleons.  The result is presented in
Fig.~\ref{fig1}. The symmetry previously discussed between $N
\leftrightarrow N_{max}+1-N$ for non-zero energy eigenstates is
obvious from the picture.

Zero energy states do not exist for $4$, $5$ and $6$ pairs as
discussed in Section II.C. From an examination of the quasispin
limit, however, one can see that the zero energy states for $1$
pair and $2$ pairs are two fold degenerate. Because the quasispin
quantum numbers corresponding to the three levels mentioned above
are $\Omega_{1/2}/2=1/2$, $\Omega_{3/2}/2=1$ and
$\Omega_{5/2}/2=3/2$ and adding these three angular momenta yields
the $J=3,2,2,1,1,0$ representations. The two fold degeneracies of
$J=1,2$ representations lead to the two fold degeneracies of zero
energy states with $1$ pair and $2$ pairs as can be seen from Eq.
(\ref{NUMBER}). We give the eigenstates of the pairing Hamiltonian
for three levels and one pair of nucleons in Table \ref{Table5}
for generic values of the $\Omega_j$ and $c_j$ parameters. The
definitions of $d_{kl}$, $a_{klm}$ and $\Gamma$ which are used in
this table are given as
\begin{equation}\label{ds}
d_{kl}=\frac{1}{|c_{j_k}|^2}-\frac{1}{|c_{j_l}|^2}~\ \ \ \ \ \ \
s_{kl}=\frac{1}{|c_{j_k}|^2}+\frac{1}{|c_{j_l}|^2}
\end{equation}
and
\begin{equation}\label{a}
a_{kml}=\Omega_{j_k}\left( d_{km}+d_{kl}\right)
+\Omega_{j_m}d_{kl}+\Omega_{j_l}d_{km}
\end{equation}
and
\begin{equation}\label{Gamma}
\Gamma =\sqrt{\left(
\Omega_{j_1}s_{23}+\Omega_{j_2}s_{13}+\Omega_{j_3}s_{12}\right)
^{2}-4\left( \Omega_{j_1}+\Omega_{j_2}+\Omega_{j_3}\right) \left(
\frac{\Omega_{j_1}}{|c_{j_2}|^2|c_{j_3}|^2}+\frac{\Omega_{j_2}}{|
c_{j_1}|^2|c_{j_3}|^2}+\frac{\Omega_{j_3}}{|c_{j_1}|^2|c_{j_2}|^2}\right)}.
\end{equation}
We see in Table \ref{Table5} that for $N=1$, we indeed have two
zero energy states. But, in Figure \ref{fig1}, we only show the
eigenvalues and not the degeneracies.

\begin{figure}[t]
  \centering
   \includegraphics{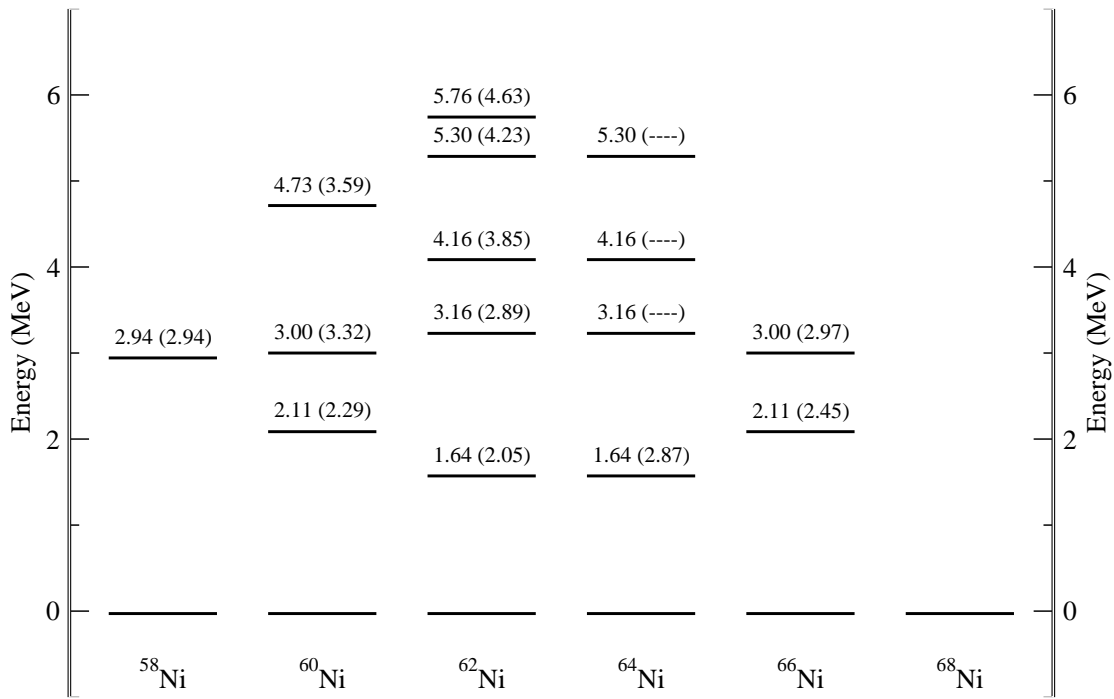} 
   \caption{Excitation spectra for Ni isotopes and comparison with experimental results
   (in brackets) \cite{auerbach}.}
   \label{fig2}
\end{figure}

\begin{table}
\begin{tabular}{|c|c|}
\hline &  \\ %
 {\bf Energy/$-|G|$} & {\bf State}
 \\ &  \\
\hline   &  \\ %
\multirow{4}{*}{$0$} & $\displaystyle
\frac{c_{j_2}c_{j_3}}{\Omega_{j_1}}d_{23}\left( a_{123}+\Gamma
\right) S_1^{+}|0\rangle +\frac{c_{j_1}c_{j_3}}{\Omega_{j_2}}
d_{31}\left( a_{231}+ \Gamma \right)S_2^{+}|0\rangle +\frac{c_{j_1}c_{j_2}}{%
\Omega_{j_3}}d_{12}\left(a_{312}+\Gamma \right) S_3^{+}|0\rangle $

\\ &  \\ \cline{2-2} &  \\ &
$\displaystyle \frac{c_{j_2}c_{j_3}}{\Omega_{j_1}}d_{23}\left(
a_{123}-\Gamma \right) S_1^{+}|0\rangle
+\frac{c_{j_1}c_{j_3}}{\Omega_{j_2}}
d_{31}\left( a_{231}- \Gamma \right)S_2^{+}|0\rangle +\frac{c_{j_1}c_{j_2}}{%
\Omega_{j_3}}d_{12}\left(a_{312}- \Gamma \right) S_3^{+}|0\rangle
$
\\ &  \\
\hline  &  \\ $\displaystyle \Omega_{j_1} |c_{j_1}|^2+\Omega_{j_2}
|c_{j_2}|^2+\Omega_{j_3} |c_{j_3}|^2$ & $\displaystyle
c^*_{j_1}\hat{S}^+_{j_1}|0 \rangle + c^*_{j_2}\hat{S}^+_{j_2}|0
\rangle+ c^*_{j_3}\hat{S}^+_{j_3}|0 \rangle$  \\ &  \\ \hline
\end{tabular}
\caption{Energies and eigenstates for three levels and one pair of
nucleons. Here, $d_{kl}$, $a_{klm}$ and $\Gamma$ are defined in
Eqs. (\ref{ds}), (\ref{a}) and (\ref{Gamma}),
respectively.}\label{Table5}
\end{table}

In order to test our formalism, we produce the excitation spectra
for six Ni isotopes, $^{58-68}$Ni (see Fig.~\ref{fig2}),
corresponding to nuclei with one to six pairs of neutrons that
occupy the first nuclear $pf$ shell above the $^{56}$Ni vacuum. We
use the occupation numbers from \cite{auerbach} and fit the
pairing strength $G$ to reproduce the experimental first excited
$0^+$ state of $^{58}$Ni. The results are in very good agreement
with experiment~\cite{table}.

\section{Conclusions and Outlook}

In this paper, we showed that the Bethe ansatz technique can be
applied to the nuclear pairing Hamiltonian with separable pairing
strengths and degenerate single particle energy levels in a purely
algebraic fashion. This algebraic method reveals a symmetry
between the energy eigenstates corresponding to at most half full
shell and those corresponding to more than half full shell.
Namely, the eigenstates with $N$ pairs of nucleons where $N\leq
N_{max}/2$ and the eigenstates with $N_{max}+1-N$ pairs of
nucleons have the same energy and these states can be found by
solving the same equations of Bethe ansatz. This symmetry is
shadowed by the effects which are not considered in this paper but
nevertheless it is manifest for the pure pairing Hamiltonian.
Using this symmetry, we are now able to complete the pairing
spectra obtained by Pan {\textit{et al}} \cite{Pan:1997rw} for the
first nuclear $sd$ shell. We have also shown that the zero energy
states do not obey this symmetry because they do not exist for
more than half full shell.

Once the pairing spectra is obtained by solving the relevant
equations of Bethe ansatz, one way to account for the effects of
the nondegeneracy of single particle energy levels is to consider
the one body term as perturbation. If the separations between the
single particle energy levels within the shell are small compared
to the pairing strength $|G|$, then one obtains small corrections
to the pairing spectra. These corrections separate the degenerate
zero energy states and shift the other states.

The Bethe ansatz technique presented here is especially powerful
if the last shell of the isotope in consideration is almost empty
or almost full. Because in this case one can find the eigenstates
and the energies by solving only a few BAE's as seen in Table
\ref{Table1}. Even if the number of levels in the shell is large,
these equations contain only a few variables and therefore can be
solved conveniently.

We also would like to point out that, in the formalism developed
here, one can use different occupation probabilities for different
isotopes. Although we construct the Bethe ansatz states starting
from the empty (or fully occupied) shell and one by one creating
(or destroying) nucleon pairs, the pairing Hamiltonian is actually
number conserving. As a result, the eigenstates presented in Table
\ref{Table1} with different number of pairs can be treated as
eigenstates of different Hamiltonians with different occupation
probabilities.

In this paper, we also considered a shell with two levels in
detail. Although the two level system is of less physical
interest, it helps us to present an alternative method of solving
the equations of Bethe ansatz. This method has certain advantages
over the conventional \textit{brute force} approach. We showed
that, in this special case, the problem of solving the equations
of Bethe ansatz for zero energy eigenstates can be transformed
into the problem of finding the roots of a certain hypergeometric
polynomial. Here, the order of the polynomial is equal to the
number of pairs in consideration. This allows us to find
analytical solutions for up to $4$ pairs of nucleons. For higher
number of pairs, one can find the roots numerically using well
established algorithms. Even in this case, finding the roots of
the hypergeometric polynomial numerically is considerably easier
than solving the corresponding equations of Bethe ansatz which are
coupled and nonlinear. Roots of various hypergeometric polynomials
and their properties are studied by many groups (see for example
\cite{Zeros} and the references therein). We would like to add
that although the technique presented here works only for a shell
with two levels, it nevertheless hints to a potentially beneficial
way of approaching the Bethe ansatz equations, in general. The
extensions of the technique to shells with higher number of levels
or to the BAE's for nonzero energy states remain to be studied.

\section*{Acknowledgements}

\noindent
This  work   was supported in  part  by   the  U.S.
National Science Foundation Grant No. PHY-0555231
at the University of  Wisconsin, and  in  part by  the University
of Wisconsin Research Committee   with  funds  granted by the
Wisconsin Alumni  Research Foundation.

\appendix

\section{The Bethe Ansatz Formalism}

In this Appendix, we briefly outline the details of the
calculations leading to the results of Section II.B. Let us start
with a Bethe ansatz state of the form
\begin{equation}\label{A1}
\hat{S}^+(y_1)\hat{S}^+(y_2) \dots \hat{S}^+(y_N)|0\rangle.
\end{equation}
This state has $N$ nucleon pairs and we assume that $2\leq N\leq
N_{max}/2$. We also assume that the variables $y_k$ are all
different from each other. Using form of the Hamiltonian given in
Eq. (\ref{10}), together with Eqs. (\ref{13}) and
(\ref{22})-(\ref{24}), it is straightforward show that the action
of the pairing Hamiltonian on the state in Eq. (\ref{A1}) is
\begin{eqnarray}\label{A2}
& &\hat{H}\hat{S}^+(y_1)\hat{S}^+(y_2)\dots
\hat{S}^+(y_N)|0\rangle \\
&=& -2|G|\left(K(y_1)+\sum_{k\neq 1}^N \frac{1}{y_1-y_k}\right)
\hat{S}^+(0)\hat{S}^+(y_2)\hat{S}^+(y_3)\dots\hat{S}^+(y_N)|0\rangle
\nonumber\\
&-& 2|G|\left(K(y_2)+\sum_{k\neq 2}^N \frac{1}{y_2-y_k}\right)
\hat{S}^+(y_1)\hat{S}^+(0)\hat{S}^+(y_3)\dots\hat{S}^+(y_N)|0\rangle \nonumber \\
&-&\dots \nonumber \\
&-& 2|G|\left(K(y_N)+\sum_{k\neq N}^N\frac{1}{y_N-y_k}\right)
\hat{S}^+(y_1)\hat{S}^+(y_2)\hat{S}^+(y_3)\dots\hat{S}^+(0)|0\rangle
. \nonumber
\end{eqnarray}
Here, $K(x)$ is given by
\begin{equation}\label{A3}
K(y)= \sum_j\frac{\Omega_j/2}{1/|c_j|^2-y} .
\end{equation}

Clearly, if we set $y_k=x^{(N)}_k$, where $x^{(N)}_k$ are the
solutions of the BAE's (\ref{38}), then all the terms on the right
hand side of Eq. (\ref{A2}) vanish. This proves that the state
given by Eq. (\ref{37}) is an eigenstate of the pairing
Hamiltonian with zero energy.

Alternatively, we can set $y_1=0$ and $y_k=z^{(N)}_{k-1}$ for
$k=2,3,\dots,N$ where $z^{(N)}_{m}$ are solutions of Eqs.
(\ref{30}). Then, using Eqs. (\ref{30}) and (\ref{A3}), we can
show that all the terms in Eq. (\ref{A2}) vanish except the first
one. In other words, (\ref{A2}) reduces to
\begin{equation}\label{A4}
\hat{H}\hat{S}^+(0)\hat{S}^+(z^{(N)}_1) \dots
\hat{S}^+(z^{(N)}_{N-1})|0\rangle=
-2|G|\left(K(0)-\sum_{k=1}^{N-1} \frac{1}{z^{(N)}_k}\right)
\hat{S}^+(0)\hat{S}^+(z^{(N)}_1) \dots
\hat{S}^+(z^{(N)}_{N-1})|0\rangle .
\end{equation}
This shows us that the state (\ref{29}) is an eigenstate. Using
Eqs. (\ref{A3}) and (\ref{A4}), one can easily show the energy of
this state is given by Eq. (\ref{32}).

Let us now consider a Bethe ansatz state in the following form:
\begin{equation}\label{A5}
\hat{S}^-(y_1)\hat{S}^-(y_2)\dots\hat{S}^-(y_{N-1})|\bar{0}\rangle.
\end{equation}
Here, $|\bar{0}\rangle$ is the state representing the full shell
defined in Eq. (\ref{14}). The operators $\hat{S}^-(y)$ annihilate
$N-1$ nucleon pairs so that the state in Eq. (\ref{A5}) has
$N_{max}+1-N$ nucleon pairs. Since we assume that $N\leq
N_{max}/2$, this state corresponds to a shell which is more than
half full.  Using the form of the Hamiltonian given in Eq.
(\ref{10}), together with Eqs. (\ref{15})-(\ref{16}) and
(\ref{22})-(\ref{24}), it can be shown that the action of the
pairing Hamiltonian on the state (\ref{A5}) is
\begin{eqnarray}\label{A6}
& &\hat{H}\hat{S}^-(y_1)\hat{S}^-(y_2) \dots
\hat{S}^-(y_{N-1})|\bar{0}\rangle
\\ &=& 2|G|\left(-K(0)+\sum_{k=1}^{N-1} \frac{1}{y_k}\right)
\hat{S}^-(y_1)\hat{S}^-(y_2)\dots\hat{S}^-(y_{N-1})|\bar{0}\rangle
\nonumber \\
&-& 2|G|\left(K(y_1)+\frac{1}{y_1}+\sum_{k\neq 1}^{N-1}
\frac{1}{y_1-y_k}\right) \hat{S}^-(0)\hat{S}^-(y_2)\dots
\hat{S}^-(y_{N-1})|\bar{0}\rangle \nonumber \\
&-& 2|G|\left(K(y_2)+\frac{1}{y_2}+\sum_{k\neq 2}^{N-1}
\frac{1}{y_2-y_k}\right)
\hat{S}^-(y_1)\hat{S}^-(0)\dots\hat{S}^-(y_{N-1})|\bar{0}\rangle
\nonumber \\
&-&\dots \nonumber \\
&-&2|G|\left(K(y_{N-1})+\frac{1}{y_{N-1}}+\sum_{k\neq
N-1}^{N-1}\frac{1}{y_{N-1}-y_k}\right)
\hat{S}^-(y_1)\hat{S}^-(y_2)\dots\hat{S}^-(0)|\bar{0}\rangle .
\nonumber
\end{eqnarray}
Here, $K(y)$ is given by Eq. (\ref{A3}). If we set
$y_k=z^{(N)}_k$, where $z^{(N)}_k$ are the solutions of the BAE's
(\ref{30}), then the state (\ref{A5}) becomes the state in Eq.
(\ref{31}). In this case, using Eqs. (\ref{30}) and (\ref{A3}), we
can show that Eq. (\ref{A6}) reduces to
\begin{equation}\label{A7}
\hat{H}\hat{S}^-(z_1^{(N)})\hat{S}^-(z_2^{(N)})\dots\hat{S}^-(z_{N-1}^{(N)})|\bar{0}\rangle=
2|G|\left(-K(0)+\sum_{k=1}^{N-1} \frac{1}{z^{(N)}_k}\right)
\hat{S}^-(z_1^{(N)})\hat{S}^-(z_2^{(N)})\dots\hat{S}^-(z_{N-1}^{(N)})|\bar{0}\rangle.
\end{equation}
This tells us that the state (\ref{31}) is also an eigenstate of
the pairing Hamiltonian. Its energy can be shown to be as in Eq.
(\ref{32}), using Eqs. (\ref{A7}) and (\ref{A3}).

\end{document}